\documentclass[a4paper,11pt]{article}
\pdfoutput=1 

\usepackage{jcappub} 

\usepackage{graphicx}
\usepackage{amsmath}
\usepackage{amssymb}
\usepackage{amsfonts}
\usepackage{color}
\newcommand{\be}{\begin{equation}}
\newcommand{\ee}{\end{equation}}
\newcommand{\nn}{\nonumber}
\def\bea{\begin{eqnarray}}
\def\eea{\end{eqnarray}}

\definecolor{darkgreen}{rgb}{0,0.6,0}

\title{\boldmath Analytic expressions for the background evolution of massive neutrinos and dark matter particles}

\author[a]{Rub\'{e}n Arjona,}
\author[b]{Wilmar Cardona,}
\author[a,1]{Savvas Nesseris,\note{Corresponding author.}}

\affiliation[a]{Instituto de F\'isica Te\'orica UAM-CSIC, Universidad Auton\'oma de Madrid,
Cantoblanco, 28049 Madrid, Spain}
\affiliation[b]{Departamento de F\'isica, Universidad del Valle, A. A. 25360 Cali, Colombia}

\emailAdd{ruben.arjona@uam.es}
\emailAdd{wilmar.cardona@correounivalle.edu.co}
\emailAdd{savvas.nesseris@csic.es}

\abstract{We provide exact analytic expressions for the density, pressure, average number density and pseudo-pressure for massive neutrinos and generic dark matter particles, both fermions and bosons. We then focus on massive neutrinos and we compare our analytic expressions with the numerical implementation in the CLASS Boltzmann code. We find that our modifications including the exact analytic expressions are in agreement to better than $10^{-4}\%$ with the default CLASS implementation in the estimation of the CMB power spectrum; our modifications do not have an impact on the performance of the code. We also provide several specific limits of our expressions at the relativistic regime, but also at late times for the neutrino equation of state.}

\begin{document}
\maketitle
\flushbottom

\section{Introduction}
\label{section:introduction}

Over the past decades Dark Matter (DM) has become a fundamental ingredient in the standard model of cosmology \cite{RevModPhys.90.045002}. Although we know relatively little about its nature, it is clear that taking into consideration DM when modelling the Universe makes it possible to explain a wide variety of astrophysical observations \cite{Ade:2015xua,Aghanim:2018eyx,Abbott:2017wau,Abbott:2018wzc}. Nowadays, it is commonly believed that DM comprises beyond Standard Model particles which move slowly with respect to the speed of light and whose interaction with other particles does not go beyond gravity: the so-called Cold Dark Matter (CDM). However, in the Standard Model of particle physics there exist candidates with similar properties which can account for a fraction of the DM in the Universe. Neutrinos weakly interact with other particles and their speed of propagation is different at late- and early-times in the cosmic evolution: in the beginning their speed of propagation is very close to the speed of light and recently they became non-relativistic. This sort of DM is usually dubbed non-Cold Dark Matter (nCDM).

Even though there is compelling evidence for flavour neutrino oscillations which implies that neutrinos are massive particles \cite{Maltoni:2004ei,Fogli:2005cq,Esteban:2018azc,deSalas:2017kay,Capozzi:2018ubv,Lesgourgues:2006nd}, current constraints do not fully determine their absolute mass scale \cite{Lesgourgues:2014zoa,Lattanzi:2017ubx,Abazajian:2016hbv}. Nevertheless, this situation is expected to change with upcoming galaxy surveys which will be able to measure the galaxy distribution on scales comparable to the horizon \cite{PhysRevD.97.123544}. Since massive neutrinos suppress power on small scales \cite{PhysRevLett.80.5255}, accurate measurements of the matter power spectrum will lead to a detection of their absolute mass scale thus reducing our ignorance of the abundance of DM in the Universe \cite{Eisenstein:1998hr,PhysRevD.70.045016,PhysRevD.97.123544}. Furthermore, measurement of neutrino masses could give hints about new fundamental theories having the Standard model of particle physics as a low-energy limit.

Due to their weakly interacting nature, neutrinos obey a collisionless Boltzmann equation. However, since neutrinos are massive particles the evolution of their phase-space distribution function is not trivial \cite{Ma:1995ey}. In order to find the unperturbed density and pressure for neutrinos current implementations in Boltzmann solvers, such as CAMB\footnote{\url{https://camb.info/}} \cite{Lewis:1999bs} and CLASS\footnote{\url{http://class-code.net/}} \cite{2011JCAP...07..034B,2011JCAP...09..032L}, employ numerical methods. Shortcomings of the numerical approach include non-trivial weighting scheme to carry out the numerical integration, possible limited precision, increase of computing time, but more importantly  hindering the understanding of the underlying physics. In this paper we show that a careful analytical treatment of the integrals makes it possible to overcome these difficulties. We provide explicit analytical solutions for the neutrino's unperturbed density, pressure, number density, and pseudo-pressure. Our expressions agree with previous phenomenological attempts of analytical approximations\footnote{See, for instance, Ref. \cite{Kunz:2016yqy}.} and also with the fully numerical implementation of the code CLASS. We have implemented our solutions in CLASS and verified that the fully numerical approach (current implementation in CLASS) and the fully analytical approach are in very good agreement. These changes in the code leave precision and computing time unchanged.

This paper is organized as follows. Firstly, in Section \ref{section:theoretical-framework} we derive our main results, namely, analytical expressions for the background evolution of massive fermions and bosons that are either relativistic or non-relativistic at decoupling. Secondly, in order to compare with previous phenomenological attempts of analytical approximations we provide asymptotic expansions at late times for the quantities governing the neutrino background evolution in Section \ref{section:asymptotic-expansions-at-late-times}. Thirdly, in Section \ref{section:numerical-results-and-implementation-in-class} we implement our analytical expressions for massive neutrinos in the code CLASS and compare with the current numerical implementation in the code. Finally, we conclude in Section \ref{section:conclusions}.

\section{Theoretical framework}
\label{section:theoretical-framework}

In this section we will derive simple analytic expressions for several key quantities that are relevant for the background evolution of massive particles, such as the average number density $n(a)$, the density $ \rho(a)$ and pressure $P(a)$ of a particle given its phase-space distribution. For the implementation in Boltzmann codes, it is also useful to calculate the derivative of the so-called pseudo-pressure, which we denote by $psP(a)$. All of these quantities are given by the following expressions:\footnote{Note that here and in what follows, we will use natural units in which $c=\hbar=k_B=1$.}
\bea
    n(a)&=& \int d^3 p f_0(p),\\
    \rho(a)&=&\int d^3 p E(p) f_0(p),\\
    P(a)&=& \int d^3 p \frac{p^2}{3E(p)} f_0(p),\\
    psP(a)&=&\int d^3 p \frac{p^4}{3E(p)^3} f_0(p),
\eea
where $p$ is the physical momentum of the particles, $a$ is the scale factor, $E$ is the energy, while the distribution $f_0(p)$ is given by
\be
f_0(p)=\frac{g_s}{e^{\frac{E(p)}{T}}\pm 1},\label{eq:distro}
\ee
where $g_s$ is the degeneracy of the species, $T$ is the temperature of the particles and the $\pm$ corresponds to fermions/bosons respectively.\footnote{We ignore the chemical potential $\mu$ in our analysis as in all realistic particles, it is much smaller than the temperature. Moreover, current bounds on the common value of the neutrino degeneracy parameter indicate that a neutrino chemical potential can be safely neglected \cite{Mangano:2011ip,Castorina:2012md,PhysRevD.66.025015,Dolgov:2002ab,Abazajian:2002qx,Serpico:2005bc,Barger:2003rt,Cuoco:2003cu}.}

As the Universe expands and cools down, the temperature will reach the decoupling temperature $T_{D}$ and all interactions will freeze out, so that the phase space distribution of Eq.~\eqref{eq:distro} of a particle with mass $M$ will remain frozen \cite{Padmanabhanbook,Lesgourgues:2006nd}
\bea
f_0(p)&=&f_{eq}\left(p\frac{a(\eta)}{a(\eta_D)},T_D\right)=\frac{g_s}{e^{\frac{\sqrt{p^2a^2/a^2_D+M^2}}{T_D}+ 1}}.
\label{eq:distro1}
\eea
Here $\eta$ is the conformal time, $f_{eq}$ is the distribution at thermal equilibrium, the subscript $D$ denotes decoupling, $a_D\equiv a(\eta_D)$, and $a\equiv a(\eta)$. Thus, we will consider two separate cases for the distribution $f_0(p)$ at the decoupling temperature $T_{D}$:
\begin{enumerate}
  \item The particles are relativistic, with energy $E(p)\sim p$;
  \item The particles are non-relativistic, with energy $E(p)= \sqrt{p^2+M^2}$.
\end{enumerate}
Note that this will only affect the distribution $f_0(p)$ and not the energy in the integrand, which can be allowed to be time-dependent. Following, we will present the analytical expressions for the thermodynamic quantities for both fermions and bosons that are relativistic and non-relativistic in Section \ref{sec:relativistic} and Section \ref{sec:non-relativistic} respectively.

\subsection{Relativistic fermions and bosons at decoupling}
\label{sec:relativistic}

In this section we are mainly interested in massive neutrinos and we will specifically focus on them, but our results are readily applicable to other massive relics that are relativistic at decoupling. Neutrino decoupling happened at $T_{D} \sim 1 \, \textrm{MeV}$ or $z \sim 10^{10}$, so at that point neutrinos are still relativistic and their distribution can be written as
\be
f_0(p)=\frac{g_s}{e^{\frac{p}{T_\nu(a)}}+ 1}.\label{eq:distroo}
\ee
Taking into account the expansion of the Universe, we see that  the physical momentum $p$ will be redshifted and can be written in terms of the comoving momentum $Q$ as $p=Q/a$, where $a=\frac{1}{1+z}$ is the scale factor and $z$ is the redshift. After neutrino decoupling the temperature scales as $T_\nu(a)=T_{\nu,0}/a$ and $T_{\nu,0}\simeq\left(\frac{4}{11}\right)^{1/3} T_{\textrm{cmb}}$ is the neutrino temperature today with a value $T_{\nu,0}\sim 1.68 \cdot 10^{-4} \, \textrm{eV}$. Therefore, the combination $Q/T_{\nu,0}$ will be constant and does not depend on the redshift, thus the distribution is frozen.

Defining $T_{\nu,0}$ as  $T_{\nu,0}=\widetilde{T}$, the previous equations for the evolution variables can be written as
\bea
n(a)&=& \frac{4\pi g_s}{a^3}\int_0^\infty dQ~Q^2 \frac{1}{e^{\frac{Q}{\widetilde{T}}}+ 1},\label{eq:n}\\
\rho(a)&=&\frac{4\pi g_s}{a^4}\int_0^\infty dQ~Q^2  \frac{\left(Q^2+a^2 M^2\right)^{1/2}}{e^{\frac{Q}{\widetilde{T}}}+ 1},\\
P(a)&=& \frac{4\pi g_s}{3a^4}\int_0^\infty dQ~Q^4  \frac{\left(Q^2+a^2 M^2\right)^{-1/2}}{e^{\frac{Q}{\widetilde{T}}}+ 1},\\
psP(a)&=&\frac{4\pi g_s}{3a^4}\int_0^\infty dQ~Q^6  \frac{\left(Q^2+a^2 M^2\right)^{-3/2}}{e^{\frac{Q}{\widetilde{T}}}+ 1},
\eea
In the previous equations all the integrals are of the form
\be
I_{n,k}\equiv\int_0^\infty dQ~Q^n  \frac{\left(Q^2+a^2 M^2\right)^{k/2}}{e^{\frac{Q}{\widetilde{T}}}+ 1},\\
\ee
where $(n,k)$ are integers. In order to calculate $I_{n,k}$ analytically, we multiply the numerator and denominator with the term $e^{\frac{-Q}{\widetilde{T}}}$ and then we use the expansion $\frac{x}{1+x}=\sum_{\beta=1}^\infty (-1)^{\beta+1} x^\beta$ for $x\leq1$, which in our case is possible as $e^{\frac{-Q}{\widetilde{T}}}\leq1$ for all $Q\in[0,\infty)$, thus our series will always converge. Then, we have
\bea
I_{n,k}&=&\int_0^\infty dQ~Q^n  \frac{e^{\frac{-Q}{\widetilde{T}}}\left(Q^2+a^2 M^2\right)^{k/2}}{e^{\frac{-Q}{\widetilde{T}}}+ 1}\nn \\
&=&\sum_{\beta=1}^\infty (-1)^{\beta+1} \int_0^\infty dQ~Q^n\left(Q^2+a^2 M^2\right)^{k/2}e^{\frac{-\beta Q}{\widetilde{T}}}.
\eea
To solve the previous integral we use Eq.~(3.389.2) from Ref.~\cite{Gradshteyn}
\begin{equation}
\int_0^{\infty}z^{2\nu-1}\left(u^2+z^2\right)^{\alpha-1}e^{-\mu z}dz=\frac{u^{2\nu+2\alpha-2}}{2\sqrt{\pi}\Gamma(1-\alpha)}G_{1,3}^{3,1} \left(\begin{matrix} 1-\nu\\
1-\alpha-\nu, 0 , \frac{1}{2}\end{matrix} \bigg| \frac{\mu^2 u^2}{4} \right),
\end{equation}
where Re $\mu>0$, Re $\nu>0$, $|\text{arg} \, u<\frac{\pi}{2}|$ and  $ G_{p,q}^{m,n} \left(\begin{matrix}a_1, \ldots, a_n & a_{n+1}, \ldots, a_p \\ b_1, \ldots, b_m & b_{m+1}, \ldots, b_q \end{matrix} \bigg| z \right)$ is the Meijer-G function. With this expression we find that
\bea
I_{n,k}=\sum_{i=1}^\infty (-1)^{i+1} \frac{(a M)^{1+k+n}}{2\sqrt{\pi}\Gamma(-k/2)}G_{1,3}^{3,1} \left(\begin{matrix} \frac{1-n}{2}\\-\frac12(1+k+n), 0 , \frac{1}{2}\end{matrix} \bigg| \frac{x_i^2}{4} \right),\label{eq:ink}
\eea
where for convenience we have set $x_i=\frac{i a M}{\widetilde{T}}$. Next we will provide the explicit expressions for each of the key background quantities mentioned earlier.

\subsubsection{Average number density}
The average number density $n(a)$ corresponds to the parameters $(n,k)=(2,0)$, so combining Eqs.~\eqref{eq:n} and \eqref{eq:ink} gives the well known result:
\begin{equation}
   n(a)=\frac{6\pi g_s\zeta(3)}{a^3}\widetilde{T}^3.
\end{equation}

\subsubsection{The density}
The density corresponds to the parameters $(n,k)=(2,1)$ and the final result can be found to be
\bea
\rho(a)&=&g_s M^4\sum_{i=1}^{\infty}\left(-1\right)^{i}
G_{1,3}^{3,1} \left(\begin{matrix} -\frac{1}{2}\\
-2, 0 , \frac{1}{2}\end{matrix} \bigg| \frac{x_i^2}{4} \right) \nn\\
&=& 2\pi^2 g_s M^4 \sum_{i=1}^{\infty}\left(-1\right)^{i}\frac{1}{x_i^3}\left(-\frac{2}{\pi}x_i^2+3x_i K_0(x_i)+\left(x_i^2-6\right)K_1(x_i) \right),\label{eq:rho_G}
 \eea
where $K_\nu(x)=H_\nu(x)-Y_\nu(x)$ is the Struve K function, $H_\nu(x)$ is the Struve H function and $Y_\nu(x)$ the usual Bessel Y function of the second kind \cite{Abramowitz}. In the relativistic limit, where $M=0$, we find
\begin{equation}
\rho(a)=\frac{7 \pi^5 g_s}{30a^4}\widetilde{T}^4.
\end{equation}
The derivative $\frac{d \rho}{dM}(a)$, which is also useful in calculations in Boltzmann solvers, corresponds to $(n,k)=(2,-1)$ and is given by
\bea
\label{eq:rho_dm}
    \frac{d\rho(a)}{dM}&=&2g_s M^3\sum_{i=1}^{\infty}\left(-1\right)^{i+1}~G_{1,3}^{3,1} \left(\begin{matrix} -\frac{1}{2}\\-1, 0 , \frac{1}{2}\end{matrix} \bigg| \frac{x_i^2}{4} \right),\nn \\
&=&2\pi^2 g_s M^3\sum_{i=1}^{\infty}\left(-1\right)^{i}\frac{1}{x_i}\left[x_i K_0(x_i)-K_1(x_i)\right].
\eea

\subsubsection{Pressure}
The pressure corresponds to the set of parameters $(n,k)=(4,-1)$ and as a result we have
\bea
\label{eq:p_G}
P(a)&=& \frac{2g_s M^4}{3} \sum_{i=1}^{\infty}\left(-1\right)^{i+1}G_{1,3}^{3,1} \left(\begin{matrix} -\frac{3}{2}\\-2, 0 , \frac{1}{2}\end{matrix} \bigg| \frac{x_i^2}{4} \right),\nn\\
&=& \frac{2g_s\pi^2 M^4}{3} \sum_{i=1}^{\infty}\left(-1\right)^{i}\frac{(3-x_i^2)}{x_i^3}\left(-\frac{2x_i^2}{(3-x_i^2)\pi}+x_i~K_0(x_i)-2K_1(x_i)\right). \label{eq:p_G2}
\eea
In the relativistic limit, where $M=0$, we find
\begin{equation}
P(a)=\frac{7 \pi^5 g_s}{90a^4}\widetilde{T}^4=\frac{\rho(a)}{3},
\end{equation}
as expected for relativistic particles.

\subsubsection{Pseudo pressure}
The pseudo-pressure corresponds to the set of parameters $(n,k)=(6,-3)$
\bea
\label{eq:pseudo_G}
psP(a)&=&\frac{4g_s M^4}{3}\sum_{i=1}^{\infty}\left(-1\right)^{i+1}G_{1,3}^{3,1} \left(\begin{matrix} -\frac{5}{2}\\-2, 0 , \frac{1}{2}\end{matrix} \bigg| \frac{x_i^2}{4} \right),\nn \\
&=&\frac{2\pi^2 g_s M^4}{3}\sum_{i=1}^{\infty}\left(-1\right)^{i}\frac{1}{x_i^3}\left(-\frac{2}{\pi}
(x_i^2+x_i^4)-3x_i(x_i^2-1)K_0(x_i)+(x_i^4+3x_i^2-6)K_1(x_i)\right).\nn\\
\eea
In the relativistic limit, where $M=0$, we find
\begin{equation}
    psP(a)=\frac{7 \pi^5 g_s}{90a^4}\widetilde{T}^4\equiv P(a)=\frac{\rho(a)}{3}.
\end{equation}

\subsubsection{Results for massive bosons}
In the case of bosons, the analytical expressions for the background are very similar to the ones found for fermions. The only difference is the factor $(-1)^i$ that appears in the sum which has to be replaced by $(-1)$. Hence, the density and pressure are
\bea
\rho(a)&=& 2\pi^2 g_s M^4 \sum_{i=1}^{\infty}\frac{1}{x_i^3}\left(\frac{2}{\pi}x_i^2-3x_i K_0(x_i)-\left(x_i^2-6\right)K_1(x_i) \right),\\
P(a)&=&\frac{2\pi^2 g_s M^4}{3}\sum_{i=1}^{\infty}\frac{(3-x_i^2)}{x_i^3}\left(\frac{2x_i^2}{(3-x_i^2)\pi}-x_i~K_0(x_i)+2K_1(x_i)\right),
 \eea
and for the average number density we obtain the well known result
\be
n(a)=\frac{8\pi g_s}{a^3}\tilde{T}^3.
\ee
In the relativistic limit where $M=0$, we find
\begin{equation}
P(a)=\frac{4 \pi^5}{45a^4}g_s\tilde{T}^4=\frac{\rho(a)}{3}.
\end{equation}
\subsubsection{Free-streaming length}
Similarly, we can also calculate the free-streaming length, i.e., the typical distance particles travel between interactions, which is defined via \cite{Lesgourgues:2018ncw,Lesgourgues:2012uu}:
\bea
k_{FT}(t)&=&\left(\frac{4 \pi G \bar{\rho}(t)a(t)^2}{v_{th}^2(t)}\right)^{1/2},\\
\lambda_{FT}(t)&=&2\pi\frac{a(t)}{k_{FT}(t)}=2\pi \sqrt{\frac23}\frac{v_{th}}{H(t)},
\eea
where $v_{th}\equiv \frac{\langle p\rangle}{m}$ is the thermal velocity and $\langle p\rangle$ the average particle momentum. After the particles become non-relativistic we can calculate the average momentum, by using the results in the previous sections for non-relativistic massive particles, as follows
\bea
\langle p\rangle &=& \frac{\int d^3 p\;p\;f_0(p)}{\int d^3 p\;f_0(p)}\\
&=&\frac{7 \pi^4}{180 \zeta(3)}T_\nu(a)\\
&\simeq& 3.15137\;\frac{T_{\nu,0}}{a}.
\eea
Finally, we have that the free-streaming length is
\bea
\lambda_{FT}(t)&=&2\pi \sqrt{\frac23}\frac{7 \pi^4}{180 \zeta(3)}\frac{T_{\nu,0}}{m a H}\nn\\
&\simeq&8.14996\;\frac1{a H(t)/H_0}\left(\frac{\textrm{eV}}{m}\right)h^{-1} \textrm{Mpc},
\eea
which is in good agreement with the result of Ref.~\cite{Lesgourgues:2018ncw,Lesgourgues:2012uu}.

\subsection{Non-relativistic fermions and bosons at decoupling}
\label{sec:non-relativistic}

When we have massive fermions that are non-relativistic at decoupling ($M_X \gg T_D$) their distribution function after the freeze out or decoupling can be written as \cite{Padmanabhanbook}
\be
f_X(p)=f_{eq}\left(p\frac{a(\eta)}{a(\eta_D)},T_D\right)=\frac{g_s}{e^{\frac{\sqrt{p^2a^2/a^2_D+M^2}}{T_D}+ 1}},\label{eq:distro1}
\ee
where the subscript $D$ denotes decoupling and $a_D\equiv a(\eta_D)$, $a\equiv a(\eta)$. Defining $\tilde{T}=T_{0}$, the comoving momentum $Q$ as $p=Q/a$ and the temperature parameter  $T_0\equiv Ta\equiv T_Da_D$, following a similar approach as in Section \ref{sec:relativistic} we can compute the average number density, the energy density and pressure as
\bea
    n(a)&=& \frac{4\pi g_s}{a^3}\int_0^\infty dQ~Q^2 \frac{1}{e^{\frac{\sqrt{Q^2+a^2_DM^2}}{\tilde{T}}}+ 1},\label{eq:n1}\\
    \rho(a)&=&\frac{4\pi g_s}{a^4}\int_0^\infty dQ~Q^2  \frac{\left(Q^2+a^2 M^2\right)^{1/2}}{e^{\frac{\sqrt{Q^2+a^2_DM^2}}{\tilde{T}}}+ 1},\label{eq:dens1}\\
    P(a)&=& \frac{4\pi g_s}{3a^4}\int_0^\infty dQ~Q^4  \frac{\left(Q^2+a^2 M^2\right)^{-1/2}}{e^{\frac{\sqrt{Q^2+a^2_DM^2}}{\tilde{T}}}+ 1}.\label{eq:press1}
\eea

\subsubsection{Average number density}
To solve Eq.(\ref{eq:n1}) we first perform a change of variables to hyperbolic functions. Then, using Eq.~(3.547.2) from Ref.~\cite{Gradshteyn} we find
\be
\label{eq:nrn}
n(a)=\frac{4\pi}{a^3}g_s a^3_DM^3\sum_{i=1}^{\infty}\left(-1\right)^{i+1}\frac{K_2(y_i)}{y_i},
\ee
where $K_n(z)$ is the modified Bessel function of the second kind and $y_i=\frac{ia_DM}{\tilde{T}}$.
\subsubsection{The density}
Using Eq.~(7.6.1) from Ref.~\cite{Moll} we find after some algebraic manipulations that the density can be written as
\be
\label{eq:nrd}
\rho(a)=\frac{4\pi g_s}{a^4}\sum_{n=0}^{\infty}\sum_{i=1}^{\infty}\left(-1\right)^{i+n}\frac{\Gamma\left(n-1/2\right)\Gamma\left(n+3/2\right)}{\pi \Gamma\left(n+1\right)}\frac{2^na_D^{n+2}M^{3-n}\tilde{T}^{n+1}}{a^{2n-1}i^{n+1}}K_{n+2}\left(y_i\right),
\ee
where $K_n(z)$ is the modified Bessel function of the second kind and $y_i=\frac{ia_DM}{\tilde{T}}$.

\subsubsection{The pressure}
Following the same procedure as with the density, we find that the pressure can be written as
\be
\label{eq:nrp}
P(a)=\frac{4\pi g_s}{3a^4}\sum_{n=0}^{\infty}\sum_{i=1}^{\infty}\left(-1\right)^{i+1}\frac{\Gamma\left(n+5/2\right)}{\Gamma\left(1/2-n\right)\Gamma\left(n+1\right)}\frac{a_DM}{\left(aM\right)^{2n+1}}\left(\frac{2a_D M\tilde{T}}{i}\right)^{n+2}K_{n+3}\left(y_i\right),
\ee
where again $K_n(z)$ is the modified Bessel function of the second kind and $y_i=\frac{ia_DM}{\tilde{T}}$.

\subsubsection{Numerical results}
We compared the analytical expressions for the average number density, density and pressure Eqs.~\eqref{eq:nrn}-\eqref{eq:nrp}, with the numerical integration of Eqs.~\eqref{eq:n1}-\eqref{eq:press1} for some realistic values of WIMP particles like neutralinos \cite{bringmann,sarkar} $z_{D} \sim 10^{13}, m_{X} \sim 25 \, \text{GeV}, T_D \sim 1 \, \text{GeV}$ and found that the agreement is better than $10^{-18}\%$ for only 4 iterations.

\subsubsection{Results for massive bosons}

In the case of bosons, the analytical expressions for the background are very similar to the ones found for fermions, the only difference being the factor $(-1)^{i+1}$ that appears in the sum which has to be replaced by $1$. Hence
\bea
n(a)&=&\frac{4\pi}{a^3}g_s a^3_DM^3\sum_{i=1}^{\infty}\frac{K_2(y_i)}{y_i},\\
\rho(a)&=&\frac{4\pi g_s}{a^4}\sum_{n=0}^{\infty}\sum_{i=1}^{\infty}\frac{\left(-1\right)^{n+1}\Gamma\left(n-1/2\right)\Gamma\left(n+3/2\right)}{\pi \Gamma\left(n+1\right)}\frac{2^na_D^{n+2}M^{3-n}\tilde{T}^{n+1}}{a^{2n-1}i^{n+1}}K_{n+2}\left(y_i\right),\\
P(a)&=&\frac{4\pi g_s}{3a^4}\sum_{n=0}^{\infty}\sum_{i=1}^{\infty}\frac{\Gamma\left(n+5/2\right)}{\Gamma\left(1/2-n\right)\Gamma\left(n+1\right)}\frac{a_DM}{\left(aM\right)^{2n+1}}\left(\frac{2a_D M\tilde{T}}{i}\right)^{n+2}K_{n+3}\left(y_i\right).\nn\\
\eea

In the non relativistic limit $M_X \gg T_D$, we can find semi-analytical expressions valid for fermions and bosons for the average number density (\ref{eq:n1}), density (\ref{eq:dens1}) and pressure (\ref{eq:press1}) in the following way
\bea
    n(a)&=&\frac{4\pi g_s}{a^3}e^{-M/T_D}\int_0^\infty dQ~Q^2e^{-\frac{Q^2}{2Ma^2_DT_D}}=\left(\frac{a_D}{a}\right)^3\left(2\pi g_s M T_D\right)^{3/2}e^{-M/T_D},\\
    \rho(a)&=&\frac{4\pi g_s}{a^4}e^{-M/T_D}\int_0^\infty dQ~Q^2  \left(Q^2+a^2 M^2\right)^{1/2}e^{-\frac{Q^2}{2Ma^2_DT_D}}\nn\\
    &=&2\pi g_s\left(\frac{a_D}{a}\right)^2M^3T_De^{\beta-M/T_D}K_1(z),\\
    P(a)&=&\frac{4\pi g_s}{3a^4}e^{-M/T_D}\int_0^\infty dQ~\frac{Q^4e^{-\frac{Q^2}{2Ma^2_DT_D}}}{\left(Q^2+a^2 M^2\right)^{1/2}}\nn \\
&=&\frac{\pi g_sM^3 e^{\beta-M/T_D}}{3}\left[MK_0(z)+(2\left(\frac{a_D}{a}\right)^2T_D-M)K_1(z)\right],
\eea
where we have to consider the series expansion $\sqrt{Q^2+a^2_DM^2} \sim a_DM+\frac{Q^2}{2a_DM}+O(Q)^4$ in the exponential and also assume that $e^{\frac{\sqrt{Q^2+a^2_DM^2}}{\tilde{T}}} \gg 1$. Again $K_n(\beta)$ is the modified Bessel function of the second kind and $\beta=\frac{a^2M}{4a^2_DT_D}$. 

\section{Asymptotic  expansions at late times}
\label{section:asymptotic-expansions-at-late-times}

The Struve K function $K_\nu(z)$ is a particular solution of the inhomogeneous Bessel differential equation
\be
\frac{d^2w}{dz^2}+\frac{1}{z}\frac{dw}{dz}+\left(1-\frac{\nu^2}{z^2}\right)w=\frac{\left(z/2\right)^{\nu-1}}{\sqrt{\pi}\Gamma\left(\nu+\frac12\right)}
\ee
and it admits the following asymptotic expansion for large values of the argument $z$ with fixed $\nu$ \cite{Abramowitz}:
\be
K_\nu(z)\sim \frac{1}{\pi}\sum_{k=0}^\infty \frac{\Gamma(k+\frac12)}{\Gamma(\nu+\frac12-k)}\left(\frac{z}{2}\right)^{\nu-2k-1},
\ee
which can be used to obtain asymptotic expansions for the quantities in the previous section. Specifically, we find
\bea
\rho(a)&=& \frac{6\pi g_s \widetilde{T}^3 M}{a^3} \left(\zeta(3)+\frac{15 \widetilde{T}^2 \zeta(5)}{2 a^2 M^2}\cdots\right),\\
\frac{d\rho(a)}{dM}&=&\frac{6\pi g_s \widetilde{T}^3}{a^3} \left(\zeta(3)-\frac{15 \widetilde{T}^2 \zeta(5)}{2 a^2 M^2}\cdots\right),\\
P(a)&=& \frac{30\pi g_s \widetilde{T}^5}{M a^5} \left(\zeta(5)-\frac{63 \widetilde{T}^2 \zeta(7)}{32 a^2 M^2}\cdots\right),\\
psP(a)&=& \frac{945\pi g_s \widetilde{T}^7}{M^3 a^7} \left(\zeta(7)-\frac{85 \widetilde{T}^2 \zeta(9)}{a^2 M^2}\cdots\right).
\eea
Keeping the zero-order terms for the density and the pressure gives an approximation at late times for the equation of state $w\equiv\frac{P}{\rho}$ as
\be
w(a)=\frac{5 \zeta(5)}{\zeta(3)} \frac{\widetilde{T}^2}{M^2} a^{-2},\label{eq:wapp}
\ee
which is accurate to a few percent at late times $z<10$. This expression is also in excellent agreement with the ansatz of Ref.~\cite{Kunz:2016yqy} that at late times the equation of state scales as $w(a)\sim 1/a^2$. Moreover, Eq.~\eqref{eq:wapp} also provides us with the exact numerical coefficient $\frac{5 \zeta(5)}{\zeta(3)} \frac{\widetilde{T}^2}{M^2}$.

\section{Numerical results and implementation in CLASS}
\label{section:numerical-results-and-implementation-in-class}

Here we present numerical comparisons between our analytic results for massive neutrinos, see Sec. \ref{sec:relativistic}, and numerical calculations of the quantities based on double precision calculations from CLASS, arbitrary precision calculations in Mathematica and the CEPHES library\footnote{\url{https://www.netlib.org/cephes/index.html}} that we used to implement the Struve K functions in C.

A lower limit on the neutrino mass of approximately $m_\nu\sim 0.06\textrm{eV}$ is settled by the existence of three-flavour oscillations (Refs.~\cite{Esteban:2018azc,deSalas:2017kay,Capozzi:2018ubv}), independently of their nature (Dirac or Majorana) and this is the value that we will use in what follows.

\begin{figure}[!t]
\centering
\includegraphics[width=0.7\textwidth]{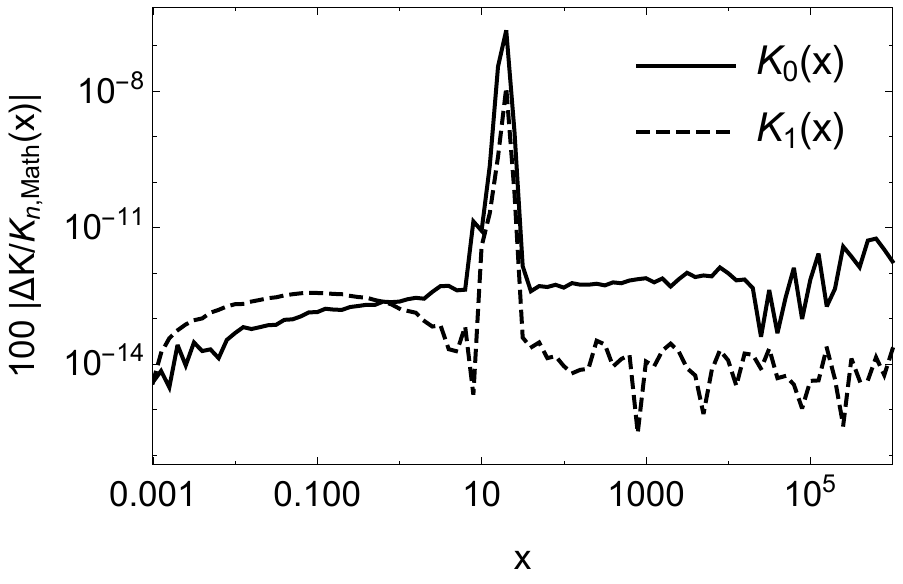}
\caption{The percent difference of the implementation in the CEPHES library vs. the arbitrary precision code of Mathematica for $K_0(x)$ (solid black line) and $K_1(x)$ (dashed black line) for $x\in[10^{-3},10^6]$. \label{fig:diffcephes}}
\end{figure}

First, we compare the implementation of the Struve K functions in CEPHES with Mathematica's arbitrary precision calculations. The results of this comparison are shown in Fig.~\ref{fig:diffcephes}, where we present the percent difference of the implementation in the CEPHES library vs. the arbitrary precision code of Mathematica for $K_0(x)$ (solid black line) and $K_1(x)$ (dashed black line) for $x\in[10^{-3},10^6]$. We find that in both cases, on average the agreement between the two codes is on the order of $\sim 10^{-12}\%$ for both functions, thus we are confident in our numerical implementation in what follows.

\begin{figure*}[!t]
\centering
\includegraphics[width=0.48\textwidth]{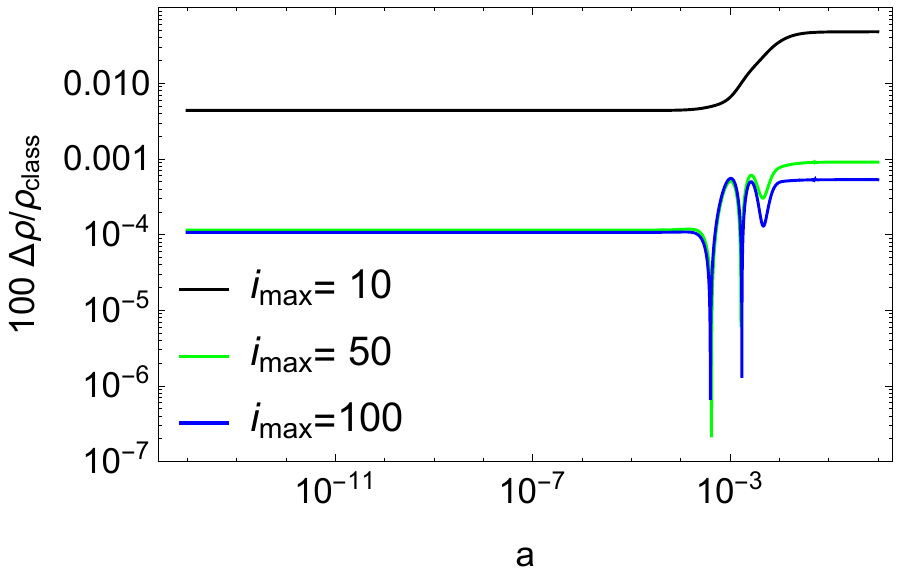}
\includegraphics[width=0.48\textwidth]{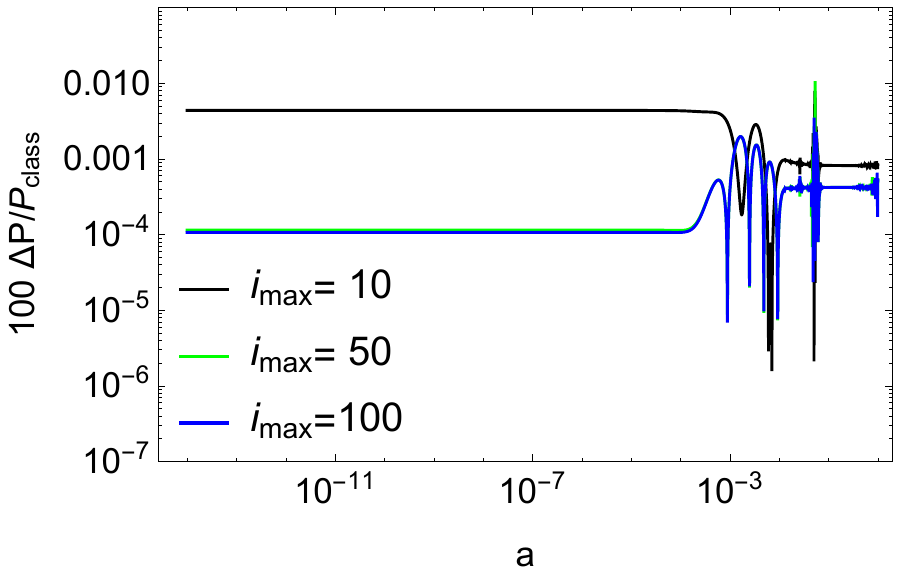}
\caption{The percent difference between the default version of CLASS and our analytical expressions for the density (left) and the pressure (right) for 10, 50 and 100 terms (black, green and blue lines). We find that keeping 50 terms in the expansion yields an accuracy of $10^{-4}\%$ on average for the density and pressure, without affecting the performance of the code. \label{fig:diff1}}
\end{figure*}

Next, we compare our numerical implementation of the analytical expressions for the neutrino density and pressure given by Eqs.~\eqref{eq:rho_G} and \eqref{eq:p_G2}, with the numerical integration done in CLASS. For this comparison we assumed no relativistic species and only 1 massive neutrino of mass $m_\nu = 0.06 \, \textrm{eV}$, while keeping all other parameters in CLASS in their default values. The results of the comparison are shown in Fig.~\ref{fig:diff1}, where we present the percent difference between the default version of CLASS and our analytical expressions for the density (left) and the pressure (right) for 10, 50 and 100 terms (black, green and blue lines) of the analytical expressions given by Eqs.~\eqref{eq:rho_G} and \eqref{eq:p_G2}. We find that keeping 50 terms in the expansion yields an accuracy of $10^{-4}\%$ on average for the density and pressure, without affecting the computational performance.

\begin{figure*}[!t]
\centering
\includegraphics[width=0.7\textwidth]{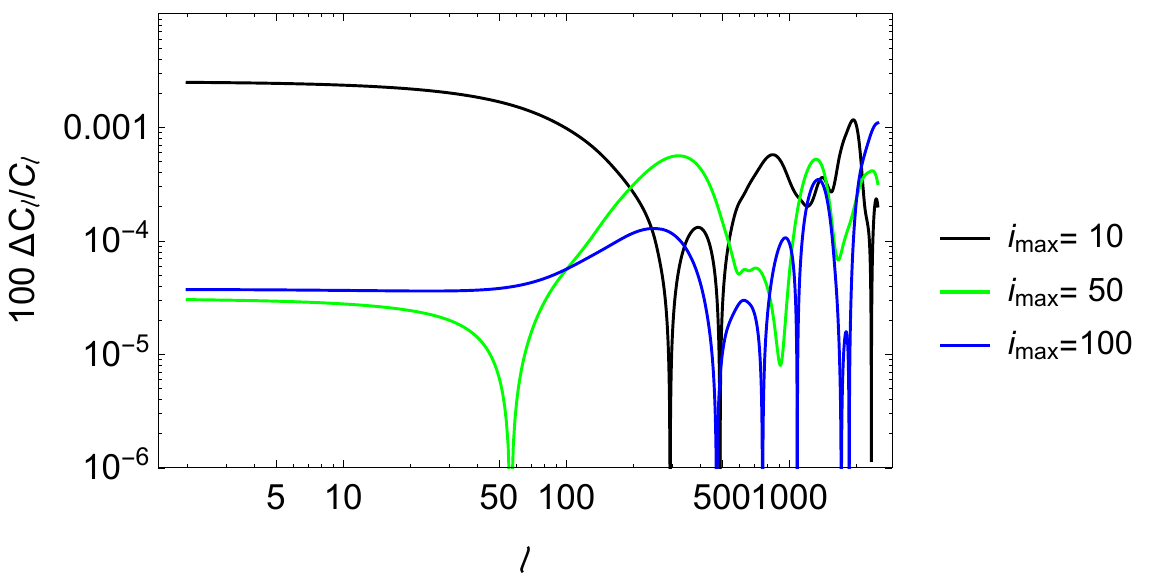}
\caption{The percent difference in the CMB power spectrum for 10 terms in the expansion (black line), 50 terms (green line) and 100 iterations (blue line). We find that keeping 50 terms in the expansion yields an accuracy of $10^{-4}\%$ on average for the CMB spectrum, without having an impact on the performance of the code. We have smoothed the data a bit to remove the oscillatory behavior at high multipoles, but this does not affect our conclusions. \label{fig:diff2}}
\end{figure*}

Then we also compare the results of the CMB power spectrum for our implementation and that of the default version of CLASS. The results of the comparison are shown in Fig.~\ref{fig:diff2}, where we present the percent difference in the CMB power spectrum for 10 terms in the expansion (black line), 50 terms (green line) and 100 iterations (blue line). We find that keeping 50 terms in the expansion yields an accuracy of $10^{-4}\%$ on average for the $C_{\ell}^{TT}$ of the CMB spectrum, without having an impact on the performance of the code.

\begin{figure*}[!t]
\centering
\includegraphics[width=0.7\textwidth]{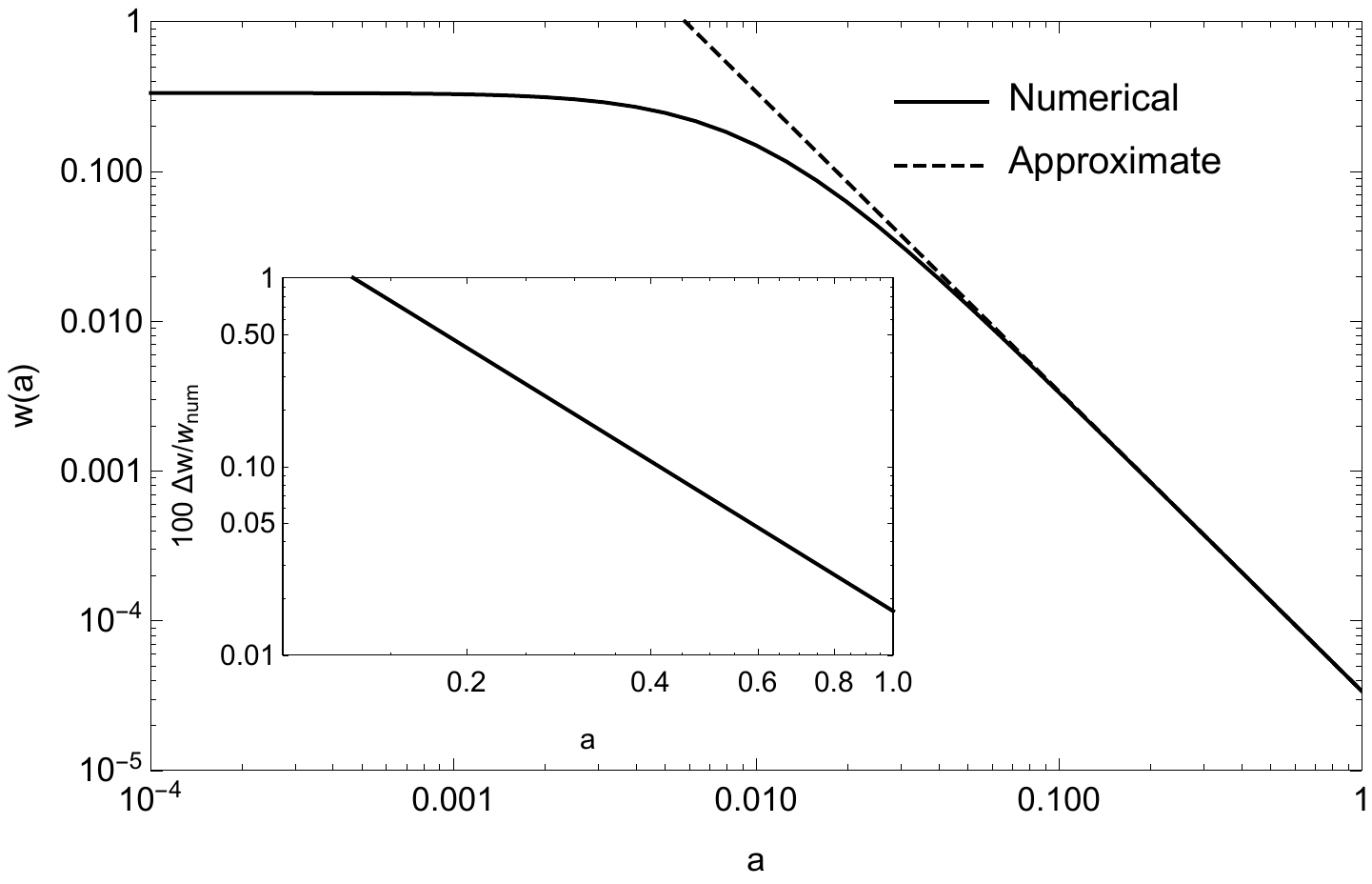}
\caption{The percent difference in the equation of state $w(a)$ between the numerical results and the approximation of $w(a)\sim a^{-2}$ given by Eq.~\eqref{eq:wapp}, for which at late times ($z<10)$, the agreement is better that $1\%$. \label{fig:diffwa}}
\end{figure*}

We also test the approximation for the equation of state $w(z)$ of the neutrinos at late times, given by Eq.~\eqref{eq:wapp}. The comparison for one massive neutrino of mass $m_\nu = 0.06 \, \textrm{eV}$ is shown in Fig.~\ref{fig:diffwa}, where we present the percent difference in the equation of state $w(a)$ between the numerical results (solid black line) and the approximation of Eq.~\eqref{eq:wapp} (dashed line), for which $w(a)\sim a^{-2}$. As can be seen in the inset plot, at late times ($z<10)$ the agreement is better that $1\%$, thus validating the ansatz of Ref.~\cite{Kunz:2016yqy}. Finally, we have also checked and confirmed that using a slightly larger neutrino mass, such as $m_\nu = 0.15-0.30 \, \textrm{eV}$, does not affect the precision of our comparison with CLASS.

\section{Conclusions}
\label{section:conclusions}

In this paper we presented simple but exact analytical expressions for the background evolution of the density $\rho(a)$, the pressure $P(a)$ and the average number density $n(a)$ for massive particles, both fermions and bosons. In both cases we considered the case when the particles are either relativistic or non-relativistic at the time of decoupling. We find that for non-relativistic massive particles the expressions are somewhat more cumbersome due to the presence of a double sum but in principle these results could be useful in future studies of dark matter candidates, such as WIMPs or any of the hypothetical superpartners of the leptons (sneutrino, etc.). 

We also specifically tested our expressions, given by Eqs.~\eqref{eq:rho_G} and \eqref{eq:p_G} for the density and pressure respectively, in the case of massive neutrinos that are still relativistic at decoupling ($z \sim 10^{10}$), assuming one neutrino with mass of $m_\nu = 0.06 \, \textrm{eV}$. We implemented our analytical expressions in the Boltzmann code CLASS and found that by keeping 50 terms in the sum, e.g., in Eqs.~\eqref{eq:rho_G} and \eqref{eq:p_G}, it is possible to achieve better than $10^{-4}\%$ accuracy with respect to the default implementation in CLASS. Our modifications in the code do not have an impact in the computational performance and avoid the involved quadrature integration scheme at the background level. Our analytical expressions provide validation for the current numerical implementations in public Boltzmann codes. By comparing CMB angular power spectra, we find the agreement between our analytical approach and the current numerical implementation is better than $10^{-4}\%$.

The main advantage of our approach is that our expressions are both exact and analytic, thus they can also provide useful intuition about the behavior of the background quantities for massive particles and how they affect the CMB. Moreover, our analytical expressions allow us to compute quantities such as the entropy density $s=\left(\rho+P\right)/T$ or the conserved number $Y=n/s$. For instance, it is possible to derive the exact behavior of the neutrino equation of state $w(a)$ at late times $(z<10)$ and show it behaves as $w\sim a^{-2}$ to better than $1\%$, in agreement with the ansatz of Ref.~\cite{Kunz:2016yqy}, thus demonstrating how fast massive neutrinos can become non-relativistic.


\textbf{Numerical Analysis Files}: The numerical codes used by the authors in the analysis of the paper and our modifications to the CLASS code will be released upon publication of the paper on the websites of the EFCLASS code \href{https://members.ift.uam-csic.es/savvas.nesseris/efclass.html}{here} and \href{https://github.com/wilmarcardonac/EFCLASS}{here}.

\acknowledgments
The authors would like to thank Juan Garc\'{\i}a-Bellido and Julien Lesgourgues for useful discussions. They also acknowledge support from the Research Projects FPA2015-68048-03-3P [MINECO-FEDER], PGC2018-094773-B-C32 and the Centro de Excelencia Severo Ochoa Program SEV-2016-0597. S.N. also acknowledges support from the Ram\'{o}n y Cajal program through Grant No. RYC-2014-15843.


\bibliographystyle{JHEP}
\bibliography{neutrinos_master}

\providecommand{\href}[2]{#2}\begingroup\raggedright\begin{thebibliography}{10}

\bibitem{RevModPhys.90.045002}
G.~Bertone and D.~Hooper, \emph{History of dark matter},
  \href{https://doi.org/10.1103/RevModPhys.90.045002}{\emph{Rev. Mod. Phys.}
  {\bfseries 90} (2018) 045002}.

\bibitem{Ade:2015xua}
{\scshape Planck} collaboration, \emph{{Planck 2015 results. XIII. Cosmological
  parameters}},
  \href{https://doi.org/10.1051/0004-6361/201525830}{\emph{Astron. Astrophys.}
  {\bfseries 594} (2016) A13}
  [\href{https://arxiv.org/abs/1502.01589}{{\ttfamily 1502.01589}}].

\bibitem{Aghanim:2018eyx}
{\scshape Planck} collaboration, \emph{{Planck 2018 results. VI. Cosmological
  parameters}},  \href{https://arxiv.org/abs/1807.06209}{{\ttfamily
  1807.06209}}.

\bibitem{Abbott:2017wau}
{\scshape DES} collaboration, \emph{{Dark Energy Survey year 1 results:
  Cosmological constraints from galaxy clustering and weak lensing}},
  \href{https://doi.org/10.1103/PhysRevD.98.043526}{\emph{Phys. Rev.}
  {\bfseries D98} (2018) 043526}
  [\href{https://arxiv.org/abs/1708.01530}{{\ttfamily 1708.01530}}].

\bibitem{Abbott:2018wzc}
{\scshape DES} collaboration, \emph{{Cosmological Constraints from Multiple
  Probes in the Dark Energy Survey}},
  \href{https://doi.org/10.1103/PhysRevLett.122.171301}{\emph{Phys. Rev. Lett.}
  {\bfseries 122} (2019) 171301}
  [\href{https://arxiv.org/abs/1811.02375}{{\ttfamily 1811.02375}}].

\bibitem{Maltoni:2004ei}
M.~Maltoni, T.~Schwetz, M.~A. Tortola and J.~W.~F. Valle, \emph{{Status of
  global fits to neutrino oscillations}},
  \href{https://doi.org/10.1088/1367-2630/6/1/122}{\emph{New J. Phys.}
  {\bfseries 6} (2004) 122}
  [\href{https://arxiv.org/abs/hep-ph/0405172}{{\ttfamily hep-ph/0405172}}].

\bibitem{Fogli:2005cq}
G.~L. Fogli, E.~Lisi, A.~Marrone and A.~Palazzo, \emph{{Global analysis of
  three-flavor neutrino masses and mixings}},
  \href{https://doi.org/10.1016/j.ppnp.2005.08.002}{\emph{Prog. Part. Nucl.
  Phys.} {\bfseries 57} (2006) 742}
  [\href{https://arxiv.org/abs/hep-ph/0506083}{{\ttfamily hep-ph/0506083}}].

\bibitem{Esteban:2018azc}
I.~Esteban, M.~C. Gonzalez-Garcia, A.~Hernandez-Cabezudo, M.~Maltoni and
  T.~Schwetz, \emph{{Global analysis of three-flavour neutrino oscillations:
  synergies and tensions in the determination of $\theta_{23}, \delta_{CP}$,
  and the mass ordering}},
  \href{https://doi.org/10.1007/JHEP01(2019)106}{\emph{JHEP} {\bfseries 01}
  (2019) 106} [\href{https://arxiv.org/abs/1811.05487}{{\ttfamily
  1811.05487}}].

\bibitem{deSalas:2017kay}
P.~F. de~Salas, D.~V. Forero, C.~A. Ternes, M.~Tortola and J.~W.~F. Valle,
  \emph{{Status of neutrino oscillations 2018: 3$\sigma$ hint for normal mass
  ordering and improved CP sensitivity}},
  \href{https://doi.org/10.1016/j.physletb.2018.06.019}{\emph{Phys. Lett.}
  {\bfseries B782} (2018) 633}
  [\href{https://arxiv.org/abs/1708.01186}{{\ttfamily 1708.01186}}].

\bibitem{Capozzi:2018ubv}
F.~Capozzi, E.~Lisi, A.~Marrone and A.~Palazzo, \emph{{Current unknowns in the
  three neutrino framework}},
  \href{https://doi.org/10.1016/j.ppnp.2018.05.005}{\emph{Prog. Part. Nucl.
  Phys.} {\bfseries 102} (2018) 48}
  [\href{https://arxiv.org/abs/1804.09678}{{\ttfamily 1804.09678}}].

\bibitem{Lesgourgues:2006nd}
J.~Lesgourgues and S.~Pastor, \emph{{Massive neutrinos and cosmology}},
  \href{https://doi.org/10.1016/j.physrep.2006.04.001}{\emph{Phys. Rept.}
  {\bfseries 429} (2006) 307}
  [\href{https://arxiv.org/abs/astro-ph/0603494}{{\ttfamily
  astro-ph/0603494}}].

\bibitem{Lesgourgues:2014zoa}
J.~Lesgourgues and S.~Pastor, \emph{{Neutrino cosmology and Planck}},
  \href{https://doi.org/10.1088/1367-2630/16/6/065002}{\emph{New J. Phys.}
  {\bfseries 16} (2014) 065002}
  [\href{https://arxiv.org/abs/1404.1740}{{\ttfamily 1404.1740}}].

\bibitem{Lattanzi:2017ubx}
M.~Lattanzi and M.~Gerbino, \emph{{Status of neutrino properties and future
  prospects - Cosmological and astrophysical constraints}},
  \href{https://doi.org/10.3389/fphy.2017.00070}{\emph{Front.in Phys.}
  {\bfseries 5} (2018) 70} [\href{https://arxiv.org/abs/1712.07109}{{\ttfamily
  1712.07109}}].

\bibitem{Abazajian:2016hbv}
K.~N. Abazajian and M.~Kaplinghat, \emph{{Neutrino Physics from the Cosmic
  Microwave Background and Large-Scale Structure}},
  \href{https://doi.org/10.1146/annurev-nucl-102014-021908}{\emph{Ann. Rev.
  Nucl. Part. Sci.} {\bfseries 66} (2016) 401}.

\bibitem{PhysRevD.97.123544}
S.~Mishra-Sharma, D.~Alonso and J.~Dunkley, \emph{Neutrino masses and
  beyond-$\mathrm{\ensuremath{\Lambda}}\mathrm{CDM}$ cosmology with lsst and
  future cmb experiments},
  \href{https://doi.org/10.1103/PhysRevD.97.123544}{\emph{Phys. Rev. D}
  {\bfseries 97} (2018) 123544}.

\bibitem{PhysRevLett.80.5255}
W.~Hu, D.~J. Eisenstein and M.~Tegmark, \emph{Weighing neutrinos with galaxy
  surveys}, \href{https://doi.org/10.1103/PhysRevLett.80.5255}{\emph{Phys. Rev.
  Lett.} {\bfseries 80} (1998) 5255}.

\bibitem{Eisenstein:1998hr}
D.~J. Eisenstein, W.~Hu and M.~Tegmark, \emph{{Cosmic complementarity: Joint
  parameter estimation from CMB experiments and redshift surveys}},
  \href{https://doi.org/10.1086/307261}{\emph{Astrophys. J.} {\bfseries 518}
  (1999) 2} [\href{https://arxiv.org/abs/astro-ph/9807130}{{\ttfamily
  astro-ph/9807130}}].

\bibitem{PhysRevD.70.045016}
J.~Lesgourgues, S.~Pastor and L.~Perotto, \emph{Probing neutrino masses with
  future galaxy redshift surveys},
  \href{https://doi.org/10.1103/PhysRevD.70.045016}{\emph{Phys. Rev. D}
  {\bfseries 70} (2004) 045016}.

\bibitem{Ma:1995ey}
C.-P. Ma and E.~Bertschinger, \emph{{Cosmological perturbation theory in the
  synchronous and conformal Newtonian gauges}},
  \href{https://doi.org/10.1086/176550}{\emph{Astrophys. J.} {\bfseries 455}
  (1995) 7} [\href{https://arxiv.org/abs/astro-ph/9506072}{{\ttfamily
  astro-ph/9506072}}].

\bibitem{Lewis:1999bs}
A.~Lewis, A.~Challinor and A.~Lasenby, \emph{{Efficient computation of CMB
  anisotropies in closed FRW models}},
  \href{https://doi.org/10.1086/309179}{\emph{Astrophys. J.} {\bfseries 538}
  (2000) 473} [\href{https://arxiv.org/abs/astro-ph/9911177}{{\ttfamily
  astro-ph/9911177}}].

\bibitem{2011JCAP...07..034B}
D.~{Blas}, J.~{Lesgourgues} and T.~{Tram}, \emph{{The Cosmic Linear Anisotropy
  Solving System (CLASS). Part II: Approximation schemes}},
  \href{https://doi.org/10.1088/1475-7516/2011/07/034}{\emph{Journal of
  Cosmology and Astro-Particle Physics} {\bfseries 2011} (2011) 034}
  [\href{https://arxiv.org/abs/1104.2933}{{\ttfamily 1104.2933}}].

\bibitem{2011JCAP...09..032L}
J.~{Lesgourgues} and T.~{Tram}, \emph{{The Cosmic Linear Anisotropy Solving
  System (CLASS) IV: efficient implementation of non-cold relics}},
  \href{https://doi.org/10.1088/1475-7516/2011/09/032}{\emph{Journal of
  Cosmology and Astro-Particle Physics} {\bfseries 2011} (2011) 032}
  [\href{https://arxiv.org/abs/1104.2935}{{\ttfamily 1104.2935}}].

\bibitem{Kunz:2016yqy}
M.~Kunz, S.~Nesseris and I.~Sawicki, \emph{{Constraints on dark-matter
  properties from large-scale structure}},
  \href{https://doi.org/10.1103/PhysRevD.94.023510}{\emph{Phys. Rev.}
  {\bfseries D94} (2016) 023510}
  [\href{https://arxiv.org/abs/1604.05701}{{\ttfamily 1604.05701}}].

\bibitem{Mangano:2011ip}
G.~Mangano, G.~Miele, S.~Pastor, O.~Pisanti and S.~Sarikas, \emph{{Updated BBN
  bounds on the cosmological lepton asymmetry for non-zero $\theta_{13}$}},
  \href{https://doi.org/10.1016/j.physletb.2012.01.015}{\emph{Phys. Lett.}
  {\bfseries B708} (2012) 1} [\href{https://arxiv.org/abs/1110.4335}{{\ttfamily
  1110.4335}}].

\bibitem{Castorina:2012md}
E.~Castorina, U.~Franca, M.~Lattanzi, J.~Lesgourgues, G.~Mangano, A.~Melchiorri
  et~al., \emph{{Cosmological lepton asymmetry with a nonzero mixing angle
  $\theta_{13}$}},
  \href{https://doi.org/10.1103/PhysRevD.86.023517}{\emph{Phys. Rev.}
  {\bfseries D86} (2012) 023517}
  [\href{https://arxiv.org/abs/1204.2510}{{\ttfamily 1204.2510}}].

\bibitem{PhysRevD.66.025015}
Y.~Y.~Y. Wong, \emph{Analytical treatment of neutrino asymmetry equilibration
  from flavor oscillations in the early universe},
  \href{https://doi.org/10.1103/PhysRevD.66.025015}{\emph{Phys. Rev. D}
  {\bfseries 66} (2002) 025015}.

\bibitem{Dolgov:2002ab}
A.~D. Dolgov, S.~H. Hansen, S.~Pastor, S.~T. Petcov, G.~G. Raffelt and D.~V.
  Semikoz, \emph{{Cosmological bounds on neutrino degeneracy improved by flavor
  oscillations}},
  \href{https://doi.org/10.1016/S0550-3213(02)00274-2}{\emph{Nucl. Phys.}
  {\bfseries B632} (2002) 363}
  [\href{https://arxiv.org/abs/hep-ph/0201287}{{\ttfamily hep-ph/0201287}}].

\bibitem{Abazajian:2002qx}
K.~N. Abazajian, J.~F. Beacom and N.~F. Bell, \emph{{Stringent constraints on
  cosmological neutrino anti-neutrino asymmetries from synchronized flavor
  transformation}},
  \href{https://doi.org/10.1103/PhysRevD.66.013008}{\emph{Phys. Rev.}
  {\bfseries D66} (2002) 013008}
  [\href{https://arxiv.org/abs/astro-ph/0203442}{{\ttfamily
  astro-ph/0203442}}].

\bibitem{Serpico:2005bc}
P.~D. Serpico and G.~G. Raffelt, \emph{{Lepton asymmetry and primordial
  nucleosynthesis in the era of precision cosmology}},
  \href{https://doi.org/10.1103/PhysRevD.71.127301}{\emph{Phys. Rev.}
  {\bfseries D71} (2005) 127301}
  [\href{https://arxiv.org/abs/astro-ph/0506162}{{\ttfamily
  astro-ph/0506162}}].

\bibitem{Barger:2003rt}
V.~Barger, J.~P. Kneller, P.~Langacker, D.~Marfatia and G.~Steigman,
  \emph{{Hiding relativistic degrees of freedom in the early universe}},
  \href{https://doi.org/10.1016/j.physletb.2003.07.039}{\emph{Phys. Lett.}
  {\bfseries B569} (2003) 123}
  [\href{https://arxiv.org/abs/hep-ph/0306061}{{\ttfamily hep-ph/0306061}}].

\bibitem{Cuoco:2003cu}
A.~Cuoco, F.~Iocco, G.~Mangano, G.~Miele, O.~Pisanti and P.~D. Serpico,
  \emph{{Present status of primordial nucleosynthesis after WMAP: results from
  a new BBN code}}, \href{https://doi.org/10.1142/S0217751X04019548}{\emph{Int.
  J. Mod. Phys.} {\bfseries A19} (2004) 4431}
  [\href{https://arxiv.org/abs/astro-ph/0307213}{{\ttfamily
  astro-ph/0307213}}].

\bibitem{Padmanabhanbook}
T.~{Padmanabhan}, \emph{{Theoretical Astrophysics - Volume 3, Galaxies and
  Cosmology}}. Dec., 2002,
  \href{https://doi.org/10.2277/0521562422}{10.2277/0521562422}.

\bibitem{Gradshteyn}
I.~S. Gradshteyn and I.~M. Ryzhik, \emph{Table of integrals, series, and
  products}. Elsevier/Academic Press, Amsterdam, seventh~ed., 2007.

\bibitem{Abramowitz}
M.~Abramowitz and I.~A. Stegun, \emph{Handbook of mathematical functions with
  formulas, graphs, and mathematical tables}, vol.~55 of \emph{National Bureau
  of Standards Applied Mathematics Series}. For sale by the Superintendent of
  Documents, U.S. Government Printing Office, Washington, D.C., 1964.

\bibitem{Lesgourgues:2018ncw}
J.~Lesgourgues, G.~Mangano, G.~Miele and S.~Pastor, \emph{{Neutrino
  Cosmology}}. Cambridge University Press, 2013.

\bibitem{Lesgourgues:2012uu}
J.~Lesgourgues and S.~Pastor, \emph{{Neutrino mass from Cosmology}},
  \href{https://doi.org/10.1155/2012/608515}{\emph{Adv. High Energy Phys.}
  {\bfseries 2012} (2012) 608515}
  [\href{https://arxiv.org/abs/1212.6154}{{\ttfamily 1212.6154}}].

\bibitem{Moll}
V.~Moll, \emph{Special Integrals of Gradshteyn and Ryzhik: the Proofs-Volume
  II}. Chapman and Hall/CRC, 2015.

\bibitem{bringmann}
T.~Bringmann and S.~Hofmann, \emph{Thermal decoupling of wimps from first
  principles}, {\emph{Journal of Cosmology and Astroparticle Physics}
  {\bfseries 2007} (2007) 016}.

\bibitem{sarkar}
A.~Sarkar, S.~Das and S.~K. Sethi, \emph{How late can the dark matter form in
  our universe?}, {\emph{Journal of Cosmology and Astroparticle Physics}
  {\bfseries 2015} (2015) 004}.

\end{thebibliography}\endgroup
\end{document}